# 45 Years of Publications in *Energy Economics*: Evolution and Thematic Trends

Maria Laura Victória Marques[1]*

Ronaldo Seroa da Motta[1]

Daniel de Abreu Pereira Uhr[2]

Julia Ziero Uhr[2]

**Abstract**

The journal *Energy Economics*, a leading international peer-reviewed outlet for economic theory in the energy sector, celebrates its 45th anniversary in 2024. This article uses a bibliometric approach to present a retrospective of the journal's contributions. The study includes all publications from the journal based on data from the *Web of Science* and *Scopus* databases. The resulting sample comprises 6,563 documents covering the period from 1979 to April 2024. The annual publication rate increased by 1.02%, with an average of 40.07 citations per document. Institutions from the United States of America and China lead in the number of articles published in the journal. A co-occurrence analysis of keywords was conducted, complemented by a sub-sample analysis, to provide a detailed view of thematic development and evolution over time. Finally, the article highlights future thematic trends of the journal, offering insights for editorial guidelines and interested authors.

[1] State University of Rio de Janeiro - Rua São Francisco Xavier, 524, Maracanã, Rio de Janeiro – RJ – Cep 20550-900 – Brazil. (*) Corresponding author: marialauravictoria@outlook.com

[2] Federal University of Pelotas - R. Gomes Carneiro, 01 - Centro, Pelotas - RS, 96010-610 – Brazil.

# 1. Introduction

*Energy Economics* is one of the leading technical-scientific sources of economic theory and its interaction with the development of the energy sector. Its original, peer-reviewed contributions extend the exploration, conversion, and use of energy, commodity and energy derivatives markets, regulation and taxation, forecasting, environment and climate, international trade, development, and monetary policy (ELSEVIER, 2024).

The journal was founded in January 1979, driven by the recognition that there was a significant gap to be filled by more rigorous studies applying economic analysis to sector-specific problems from both producer and consumer perspectives (Motamen & Robinson, 1979). In October 1994, the publishing office was transferred to Elsevier Sciences Ltd., which took over the publication, editing, production, and reprinting of *Energy Economics* under the direction of editor Derek Bunn. Since then, the journal has grown in the number of publications and impact factor, ranking in the first quartile of Economics, Econometrics, and Energy journals according to the latest *SCImago Journal Rank* (SJR) report. The editorial team comprises one editor-in-chief (Richard S. J. Tol), nine co-editors, 42 assistant editors, and one honorary editor (J. P. Weyant).

*Energy Economics* celebrates its 45th anniversary in 2024. This article aims to conduct a historical review of the journal, highlighting its key bibliographic aspects. A bibliometric method will be applied to achieve it, covering the entire period of its activities. This study will be based on three bibliometric procedures: co-citation analysis, co-authorship analysis, and keyword co-occurrence analysis. Co-authorship analysis allows us to assess the degree of collaboration among researchers, institutions, and countries. Co-citation analysis reveals the structure of the scientific domain based on the frequency with which two authors are cited together by another work (Small, 1973). Finally, keyword co-occurrence analysis helps identify the directions in which the research field has developed and its future trajectory.

The remainder of this article is structured as follows: Section 2 presents a historical overview of *Energy Economics*. Section 3 presents the method and data. Section 4 provides the main descriptive results, including co-citation, co-authorship and co-occurrence analyses. Section 4 offers a thematic overview of the journal's most established fields of exploration and discusses trends and *hotspots* for *Energy Economics*. Finally, Section 5 highlights the main findings and conclusions.

# 2. Historical Overview of *Energy Economics*

The energy sector has been central to global economic and geopolitical development over the centuries, but the scrutiny and analysis it has received in recent decades is particularly noteworthy. The 1973 oil crisis, marked by an embargo that drastically reduced oil supply and led to price shocks, awakened unprecedented awareness of global energy vulnerability and the need for deeper studies on energy economics. This historical context generated a substantial body of literature, initially more journalistic than analytical, focused on addressing the immediate emergencies imposed by such crises.

As a dedicated forum for specialized studies, *Energy Economics* emerged as a direct response to the fragmentation and dispersion of energy studies, which had been scattered across various economic journals without a specific focus on applying economic theory to the energy sector (Motamen & Robinson, 1979). The journal aimed to address this gap by integrating and advancing theoretical and applied research in energy economics. Its primary goal was to address the perspectives of both producers and consumers in the sector, understanding that the absence of this comprehensive view often leads to forecasts and policies based on erroneous or poorly informed assumptions.

Throughout its history, the journal has included numerous renowned economists, some of whom have received prestigious honors. Joseph Stiglitz was awarded the Sveriges Riksbank Prize in Economic Sciences in Memory of Alfred Nobel in 2001 and was part of the initial editorial board in 1979. The following year, William Nordhaus joined the editorial board. Nordhaus, a prominent economist in Environmental Economics and Climate Change, was recognized with the Nobel Prize in Economics for integrating climate change elements into long-term dynamic macroeconomic models[3]. Other distinguished economists who have served on the editorial board of *Energy Economics* include David W. Pearce, Robert Pindyck, Geoffrey Heal, Karen Fisher-Vanden, David Popp, Reyer Gerlagh, Richard Newell, and Richard Green, the latter group being highly influential in the field of Energy Economics. The contributions of these prominent researchers have strengthened the journal's global impact on discussions and public policies related to its area of specialization.

The presence of long-serving editors-in-chief and smooth transitions between them has characterized the editorial leadership of *Energy Economics*. Table 1 describes the time intervals during which each editor-in-chief (and co-editors, when applicable) led the journal's editorial decisions. Homa Motamen served as the coordinating editor of *Energy Economics* from 1979 to 1993, a period of 14 years, with Colin Robinson participation from 1979 to 1982. Richard Tol has been the editor-in-chief from 2004 to the present, a period of at least 20 years. From 2004 to 2007, he co-edited with John Weyant, from 2008 to 2013 with Beng W., and from 2014 to 2015 with U. Soytas. Since 2016, Richard Tol has served as the sole editor-in-chief.

**Table 1 – Editorial Board (1979 – 2024)**

| Year | Editor in chef | #EBM |
|---|---|---|
| 1979 - 1982 | Homa Motamen; Colin Robinson | 26 |
| 1983 - 1993 | Homa Motamen | 28 |
| 1996 - 1999 | Derek Bunn | 25 |
| 2000 - 2003 | Derek Bunn; Musa Essayyad; Richard Tol | 33 |
| 2004 – 2007 | Richard Tol; John Weyant (Editor Honorário: Derek Bunn) | 34 |
| 2008 - 2013 | Beng W.; Richard Tol; John Weyant | 34 |
| 2014 - 2015 | Beng W.; Richard Tol; John Weyant; U. Soytas | 36 |
| 2016 – 2024* | Richard Tol (Editor Honorário: John Weyant) | 40 |

**Source:** Authors' elaboration. (*) April 2024.

A second notable point is the honorary editor policy applied to the journal. Derek Bunn guided the editorial policies of *Energy Economics* from 1996 to 2003. From 2004 to 2006, Bunn was listed as an honorary editor. Similarly, John Weyant, who co-edited from 2007 to 2014, has been listed as an honorary editor since 2017. This recognition of past editors honors their contributions and helps consolidate and mature the journal's literature, avoiding disruptive discontinuities that could hinder its accumulated progress.

---

[3]Paul Romer also received recognition in the same year.

Another significant trend is the increasing number of members on the editorial board. From 1979 to 1997, the editorial board of *Energy Economics* comprised around 27 members. This number grew to 35 between 2002 and 2016 and 40 from 2017 onwards. As of March 2024, the editorial board comprised 53 members. Figure 1 illustrates the annual publication volume alongside the number of editorial board members, demonstrating this growth.

**Figure 1 – Annual publications and number of members on the editorial board**

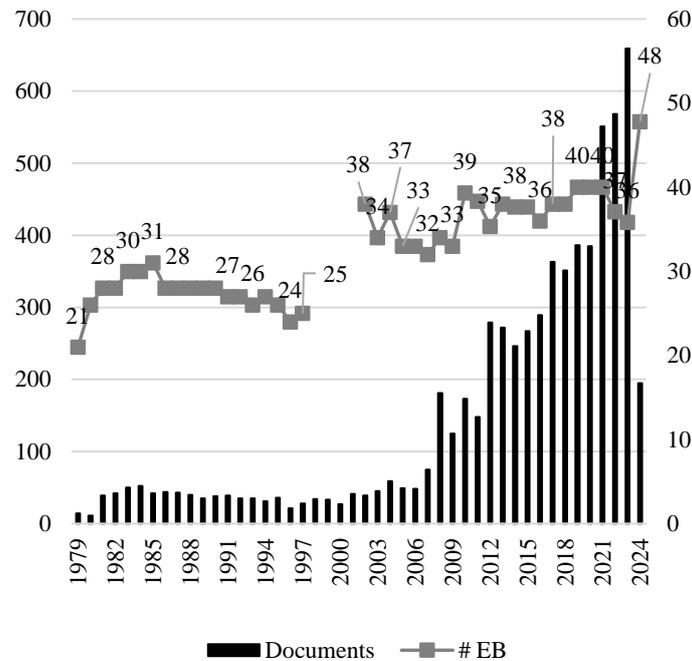

**Source:** Authors' elaboration with data from *Web of Science* and *Scopus*.

*Energy Economics* relocated its publication, editing, production, and reprinting office to Elsevier Sciences Ltd[4] in the United Kingdom in 1994. Motivated by technological advancements, the same year saw the publication of an editorial letter encouraging authors to submit their manuscripts on diskette in addition to a printed copy. This shift to digital submissions aimed to maintain text integrity, reduce compositional errors, and shorten review times. The journal streamlined the production process by editing and encoding directly from the disk, ultimately enhancing productivity. The significant increase in publications observed from 2007 onwards can be attributed mainly to these digital tools in the submission, editing, and publication processes, facilitating more frequent publications and contributing to the journal's more significant impact and internationalization.

The establishment of *Energy Economics* reflects a broader recognition of the economic relevance of the energy sector, not only as an academic field but also as a critical area for policy and business decisions. Thus, the journal has played a pivotal role in consolidating energy economics as a distinct field within the broader economic discipline. It has also provided valuable insights for policymakers and business practitioners, shaping the future of global energy. The journal continues to reflect on the impacts and lessons from historical episodes such as the 1973 oil crisis while addressing future challenges.

---

[4] *Butterworth-Heinemann* is a publisher specializing in professional books covering diverse areas, such as engineering, information technology, business and healthcare. It was originally an independent entity, but became a brand of *Elsevier Science Ltd*. *Elsevier* acquired *Butterworth-Heinemann* in the late 1990s, integrating it into its extensive portfolio of academic and technical publications. *Elsevier* itself is a leading global information analytics company specializing in science and healthcare. This acquisition allowed *Elsevier* to strengthen its offering in several professional areas through *Butterworth-Heinemann's* specialized content (DBpedia, 2024; Medlik, 2016).

## 3. Method and Data

Understanding the development of *Energy Economics* is integral to comprehending and consolidating a highly relevant field of theoretical and empirical exploration within economics. To achieve this objective, we will employ the bibliometric method. Bibliometrics is an interdisciplinary scientific approach aimed at quantifying the academic output of individuals and institutions on a specific topic using the citations and co-citations of each unit of analysis as a relevance indicator (Garfield, 1979; Glänzel, 2003). Subsequently, qualitative conclusions are drawn from its graphical and statistical results (Ball, 2018). Schwert (1993) and Chen et al. (2009) emphasize that such results enable the identification of the past and evolution of journals, offering a meticulous retrospective. This method is well-established and has been used for similar purposes by other authors (Allen & Kau, 1991; Cobo et al., 2015; Hoepner et al., 2012; Laengle et al., 2017; Viglia et al., 2022; Yao et al., 2023).

The bibliometric process will be divided into four stages: data compilation, database description, and bibliometric analysis (co-authorship, co-citation, co-occurrence and thematic mapping). In the first stage, all documents published by *Energy Economics* were compiled using the following advanced query: Source Titles = ("*Energy Economics*"). This process was conducted on the *Scopus* and *Web of Science* databases. The search was performed in April 2024, when all publications from 2023 were available. Additionally, three volumes from 2024 had already been published, and others were available online.

Next, the database description and method application were performed using VOSviewer and the R bibliometric package, *Bibliometrix*, for statistical estimations and graphical representations. VOSviewer is a bibliometric analysis software based on information visualization that was developed in Java. Unlike many other software tools used for bibliometric mapping, VOSviewer focuses mainly on the visual representation of bibliometric maps. Its capabilities are especially beneficial due to its high computational power, enabling the presentation of large bibliometric maps in an easy-to-understand and detailed manner (van Eck & Waltman, 2010). The Bibliometrix package (Aria & Cuccurullo, 2017) allows the execution of thematic maps with significant flexibility in altering the underlying analysis parameters, facilitating the selection of the most suitable clustering algorithm for the sample characteristics.

For each unit of analysis (authors, documents, countries, organizations, and keywords), cluster analysis will be conducted. Cluster analysis, or clustering, is a multivariate statistical procedure to partition elements into two or more clusters based on their similarity according to predefined criteria (Santos et al., 2020). These similarities provide valuable insights into how the literature, in terms of its concepts, methods, and theories, is organized. This approach allows for analyzing connections among actors over time and across geospatial dimensions (Aria & Cuccurullo, 2017).

## 4. Results

The inaugural volume of *Energy Economics*, published in 1979, consisted of four issues containing 30 articles. Since then, the annual publication average has varied across different periods. From 1979 to 2007, the annual average was 35 articles. From 2008 to 2016, *Energy Economics* published an average of 212 articles per year. This number continued to grow, reaching 362 articles per year from 2017 to 2020 and 582 articles annually from 2020 to 2023. Between 1979 and 1998, *Energy Economics* operated with a single volume distributed quarterly. By 1999, six issues were published annually in a single volume. From 2013 to 2015, a bi-monthly volume was published. From 2016 to 2020, eight volumes were published annually. Since 2021, *Energy Economics* has been publishing monthly volumes.

A total of 6,563 documents were cataloged from 1979 to 2024, including 5,985 articles, 64 editorial materials, 32 book reviews, 152 conference papers, 13 notes, 23 reviews, 11 letters, 18 corrections, one retraction, 13 notes, and one withdrawn item. Additionally, we divided the publications into four phases. In the first stage (1979-2007), there was a slow increase from 30 to 50 articles published annually. In the second stage (2008–2016), the annual publications averaged 157 in the first four years, rising to 270 annually in the remaining period. In the third phase (2016-2020), approximately 355 documents were published yearly. In the fourth phase (2021-2024), the average increased to 593 documents, with 2023 marking the highest number of publications in *Energy Economics* history (659).

Table 2 – Citation Structure of *Energy Economics* (1979 – 2023)

| Year | ≥200 | ≥100 | ≥50 | ≥10 | ≥5 | TC | AC |
|---|---|---|---|---|---|---|---|
| 1979 – 1983 | 0 | 2 | 7 | 33 | 59 | 1435 | 11,21 |
| 1984 – 1988 | 1 | 2 | 7 | 57 | 98 | 2120 | 10,60 |
| 1989 – 1993 | 3 | 9 | 14 | 55 | 90 | 3421 | 19,66 |
| 1994 – 1998 | 4 | 13 | 40 | 107 | 126 | 6261 | 42,59 |
| 1999 – 2003 | 13 | 34 | 70 | 144 | 162 | 13378 | 73,10 |
| 2004 – 2008 | 47 | 124 | 203 | 345 | 380 | 34941 | 87,57 |
| 2009 - 2013 | 67 | 192 | 418 | 866 | 933 | 69581 | 72,10 |
| 2014 – 2018 | 60 | 200 | 476 | 1.272 | 1.426 | 75890 | 50,86 |
| 2019 – 2024 | 28 | 130 | 337 | 1.325 | 1.762 | 62716 | 24,98 |
| Total | 223 | 706 | 1.572 | 4.204 | 5.036 | 269.743 | 43,63 |

**Source:** Authors' elaboration with data from *Web of Science* and *Scopus*.

Table 2 presents the citation structure of *Energy Economics* from 1979 to 2023. The periods are divided into four-year intervals, and the values in the table refer to sums and the interval's midpoint. First, we observe that the number of citations has increased over time: overall, *Energy Economics* articles were cited 269,743 times from 1979 to April 2024. The overall average citation (AC) is 43.63. The period from 2014 to 2018 saw the highest number of citations in the journal's history. The decline in the most recent interval is primarily due to accumulative nature of citations over time — explaining the reduction in both total citations (TC) and average citations (AC). The highest average citation per article was observed from 2004 to 2008, at approximately 88 annual citations per document.

Additionally, Table 2 shows that 223 articles were cited at least 200 times, representing 3.49% of the journal's publications; 706 articles were cited at least 100 times (11.05%), 1,572 articles were cited at least 50 times (24.60%), 4,204 articles were cited at least ten times (65.80%), and 5,036 articles were cited at least five times (78.82%). Only 420 articles (6.57%) received no citations, of which 162 (2.53%) were published in 2023. It is important to note that these numbers constantly change as articles attract more attention and gain maturity through attempts at refutation or corroboration.

To assess the influence and performance of *Energy Economics*, we applied two widely used indices: the h-index and the g-index. The g-index is the top g articles receiving at least g² citations together. For a given journal, the h-index corresponds to the most significant number of h such that h articles have received at least h citations each (Hirsch, 2005; Egghe, 2006).[5] The h-index aims to overcome the limitations of other metrics that may excessively favor the number of publications or citations. The g-index, in turn, highlights the journal's ability to publish articles that attract attention (in terms of citations) and exhibit significant influence, as evidenced by the substantial volume of citations. Both indices are complementary for analyzing scientific output. Figure 2 presents the h-index and g-index of *Energy Economics* from 1979 to 2023. The graph is based on annual data, which was not yet fully available for the year 2024 - which is why it was not included in this calculation.

**Figure 2 – *g-index* e *h-index* do periódico *Energy Economics* (1979 – 2023)**

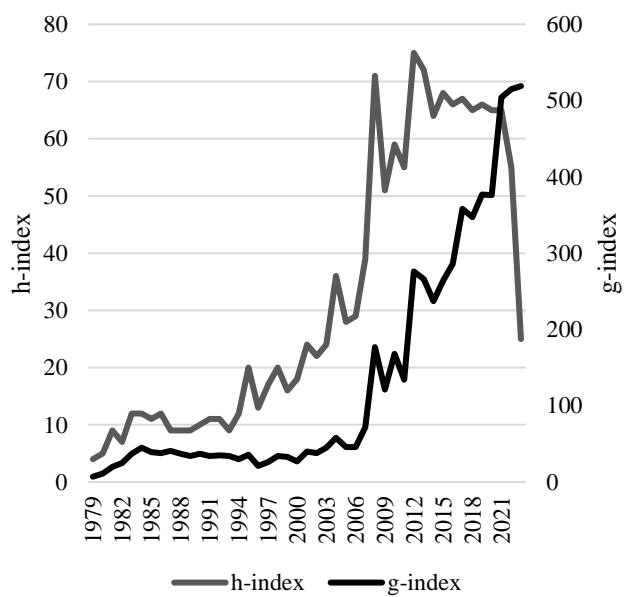

**Source:** Authors' elaboration with data from *Web of Science* and *Scopus*.

A rising trend is observed from 1997 to 2013 for the h-index and from 2006 to 2021 for the g-index. In 2012, the h-index reached its highest value (75), meaning that 75 articles 2012 received at least 75 citations. However, from 2013 to 2019, the h-index remained stable, when from 2019 onwards it showed a strong decline. The highest g-index value was observed in 2021 (504), indicating that 504 articles published by *Energy Economics* collectively received at least 254,016 citations.

Finally, to complete the analysis of the journal's performance, we present the impact factor (IF) indicator. The IF for a given year is the average number of citations received in that year by articles published in the journal during the two preceding years. A journal might have a high IF but a moderate h-index if only recent articles are highly cited. Conversely, a journal with a high h-index but a moderate IF might indicate consistent and sustained influence over time, rather than recent surges in citations. Finally, a high g-index shows that the journal has published exceptionally influential papers cumulatively considered, which might not be captured by JIF or h-index alone. Figure 3 shows the evolution of the impact factor (IF) for *Energy Economics* from 1997-2022[6].

---

[5] In g-index, g is an integer that matches the following criteria: $\sum_{i=1}^{g} c_i \geq g^2$; where g is the number o documents and $c_i$ the number of citations of the ith study. As for the h-index, h is an integer such that the following criterion is met: $h = \max\{k: c_k \geq k\}$, where k is the number of articles.

**Figure 3 – Impact Factor of *Energy Economics* (1997 – 2022)**

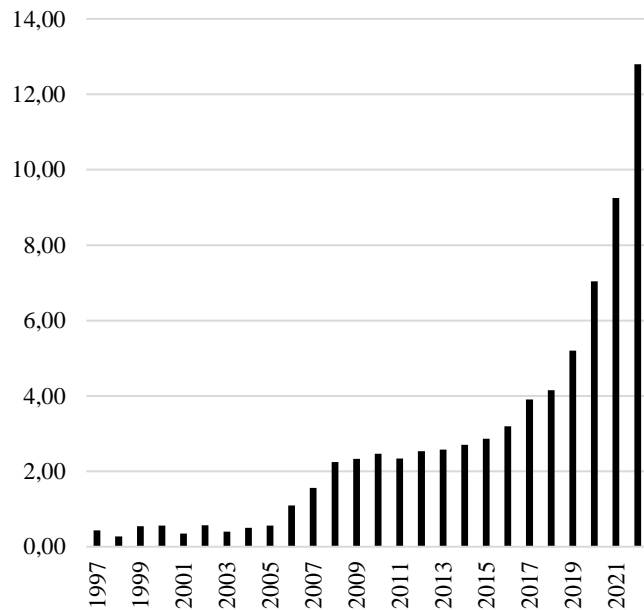

**Source:** Compiled by the authors based on *Journal Citation Reports* (*Clarivate Analytics*) statistics.

As illustrated in Figure 3, *Energy Economics* had an impact factor of 0.431 when Journal Citation Reports (JCR) statistics began in 1997. The level remained stable until 2005 when the impact factor rose from 0.564 to 1.098 in 2006. Since 2007, the impact factor of *Energy Economics* has increased, reaching 3.2 in 2016, 7.04 in 2020, and 12.8 in 2022. An ascending JIF suggests that articles published in the journal in the last two years are receiving more citations on average. This indicates a recent rise in the journal's influence and visibility within the academic community.

In recent years (2019-2023), the IF and g-index have increased while the h-index has decreased. The fact that a journal's Impact Factor (IF) and g-index are rising while the h-index is falling might be linked to a polarization in citation distribution. This suggests that a small number of very influential articles are receiving a lot of citations, which is increasing the IF and g-index. The h-index, which gauges cumulative influence generally, may be declining as a result of a decrease in the total number of consistently referenced publications. This discrepancy may also be the consequence of publication policies that have been changed to prioritize high-impact papers or a deliberate concentration on publishing ground-breaking research, which has led to less consistent citation patterns among the journal's publications. We'll look more closely at the bibliometric metrics of the various units of analysis in the sections that follow.

---

[6] Journal Citation Reports (JCR) is an annual electronic publication from Clarivate Analytics that provides information about academic journals indexed in Web of Science Core Colletcion (Clarivate, 2024). The report began to be published in 1997, and the latest year available is 2022, published in 2023.

**4.1. Most Cited Articles**

Table 3 lists the top 10 most cited articles in *Energy Economics* from 1979 to 2024. CT constitutes the total value of citations received by the article; AY consists of the number of years from the date of publication until 2024. Since the number of citations is affected by the time available for access, the column CAY (Citations per Active Year) measures the average number of citations for the years published in the 2023 account. Topping the list is the article "*Oil price shocks and stock market activity*" by Peter Sadorsky, published in 1999, with 1,040 citations. The second most cited article is by Park & Ratti (2008), titled "*Oil price shocks and stock markets in the U.S. and 13 European countries.*" Both articles delve into the impact of price shocks in the oil sector on stock markets. The articles by Sadorsky (2009) and Sadorsky (2012), which also appear in the most cited articles, explore the nexus between the energy sector and the financial market.

**Table 3 – Most Cited Articles published in *Energy Economics* between 1979 and 2024**

| TC | Title | AY | CAY | Reference |
|---|---|---|---|---|
| 1040 | Oil price shocks and stock market activity | 25 | 43,33 | Sadorsky (1999) |
| 762 | Oil price shocks and stock markets in the U.S. and 13 European countries | 16 | 50,80 | Park & Ratti (2008) |
| 716 | Do economic, financial and institutional developments matter for environmental degradation? Evidence from transitional economies | 14 | 55,08 | Tamazian & Rao (2010) |
| 648 | The relationship between energy consumption, energy prices and economic growth: time series evidence from Asian developing countries | 24 | 28,17 | Asafu-Adjaye (2000) |
| 594 | Oil price shocks and stock market activity | 15 | 42,43 | Sadorsky (2009) |
| 579 | The effect of urbanization on CO2 emissions in emerging economies | 10 | 64,33 | Sadorsky (2014) |
| 559 | CO2 emissions, energy consumption and economic growth nexus in MENA countries: Evidence from simultaneous equations models | 11 | 55,90 | Omri (2013) |
| 552 | Energy consumption and GDP in developing countries: A cointegrated panel analysis | 19 | 30,66 | Lee (2005) |
| 546 | Total factor carbon emission performance: A Malmquist index analysis | 14 | 42 | Zhou & Ang (2010) |
| 498 | Correlations and volatility spillovers | 12 | 45,27 | Sadorksy (2012) |

| | between oil prices and the stock prices of clean energy and technology companies | | | |
|---|---|---|---|---|
| 496 | Energy consumption and economic growth: Evidence from China at both aggregated and disaggregated levels | 16 | 33,06 | Yuan et al. (2008) |
| 496 | A multivariate cointegration analysis of the role of energy in the US macroeconomy | 24 | 21,56 | Stern (2000) |
| 487 | Economic growth, CO2 emissions and energy consumption: What causes what and where? | 6 | 97,40 | Acheampong (2018) |

**Source:** Authors' elaboration with data from *Web of Science* and *Scopus*.

The work of Acheampong (2018), Sadorsky (2014), Omri (2013), Tamazian & Rao (2010), and Acheampong (2018) stands out with an MCAA of 97.40, despite being also the most recently published article. This result highlights the increase in the impact factor and g-index, an analysis previously presented, due to its discrepant value in terms of average citation per active year. The work analyzes the causality between economic growth, carbon dioxide emissions, and energy consumption, a study of great relevance in energy economics and environmental economics. The topic is also addressed in the study by Sadorksy (2014), which analyzes the impact of urbanization on CO2 emission levels. CO2 emissions became a hot topic in the late 2010s due to a confluence of scientific, social, and political factors. By this time, overwhelming scientific evidence had cemented the link between CO2 emissions and climate change, with numerous studies projecting severe environmental and economic impacts if emissions were not curtailed. Public awareness and concern about climate issues had also reached unprecedented levels, influenced by high-profile climate activists, global movements, and extensive media coverage. Politically, global initiatives such as the Paris Agreement, which was adopted in 2015, underscored the urgency for collective action to reduce emissions. Technological advancements and economic shifts also played roles; the late 2010s saw significant strides in renewable energy, making sustainable alternatives more viable and challenging the long-standing dominance of fossil fuels. These dynamics collectively pushed CO2 emissions to the forefront of public discourse and policy agendas in the late 2010s, unlike in earlier years when the topic lacked a similar degree of urgency and widespread acknowledgment.

**4.2. Countries with the Highest Number of Publications in *Energy Economics***

The results related to the degree of internationalization of *Energy Economics* through the collaboration of each country and cooperation with other countries and regions in each publication are presented in this subsection. This analysis is based on the address information of the authors listed in the publications. It allows us to observe the academic power and scientific production of countries and identify priority regions for those seeking to understand the contribution of *Energy Economics* to the energy economics literature. Table 4 presents the ten most productive countries. It is important to note that a document can contribute to the count of more than one country. It occurs because the counting process considers the author's and other co-authors' countries, leading to possible double counting. The columns AC (citations per document) and TC (total citations) allow us to observe the impact of each country on the research field.

From 1973 to 2024, approximately 148 countries contributed to the journal. The countries are listed by the total number of publications (TP). Other indicators are also presented per country, including the total number of citations (TC), average citations (AC), single-country publications (SCP), number of multi-country publications (MCP), and the ratio of multi-country publications to total publications (MCP-ratio).

**Table 4 – Most Productive and Influential Countries in *Energy Economics* between 1979 and 2024**

| Country | TP | SCP | MCP | MCP-ratio | TC | AC |
|---|---|---|---|---|---|---|
| United States | 1,298 | 871 | 367 | 0,283 | 54,648 | 42,10 |
| China | 1,244 | 931 | 373 | 0,300 | 47,048 | 37,82 |
| United Kingdom | 412 | 237 | 175 | 0,425 | 16,882 | 40,98 |
| Germany | 355 | 241 | 114 | 0,321 | 14,604 | 41,14 |
| Australia | 351 | 218 | 133 | 0,379 | 13,016 | 37,08 |
| France | 212 | 143 | 69 | 0,325 | 11,026 | 52,01 |
| Canada | 206 | 146 | 60 | 0,291 | 9,235 | 44,83 |
| Spain | 202 | 127 | 75 | 0,371 | 9,193 | 45,51 |
| Italy | 169 | 111 | 58 | 0,343 | 6,964 | 41,21 |
| Norway | 124 | 101 | 23 | 0,185 | 6,139 | 49,51 |

**Source:** Authors' elaboration with data from *Web of Science* and *Scopus*.

In terms of total publications, the United States appears as the most productive country, with 1,298 publications, followed by China (1,244) and the United Kingdom (412). This composition of countries at the top is similar to other bibliometrics of specialized journals (Yao et al., 2023). Regarding citations, it can be stated that the United States, China, and the United Kingdom, in addition to the countries with the highest number of publications, are also the most influential according to bibliometric criteria (Ball, 2018). Besides these, Germany, Australia, France, and Canada make up the list. When considering the average citation per publication, Norway becomes the leading country, with an average citation of 51.18, followed by France (49.94), Italy (49.30), and the United Kingdom (46.23).

The degree of internationalization of a journal indicates its insertion and dialogue about different contexts and problems in the energy sector, allowing it to compose a solid scientific framework on its specialized topic. In Table 4, the indicators SCP, MCP, and MCP-ratio seek to measure aspects of international collaboration in *Energy Economics*. In terms of publications where all authors are affiliated with the same country, China tops the ranking (931), followed by the United States (871) and Germany (241); these values correspond to 33.62%, 67.10%, and 67.88% of their respective total productions. In terms of international collaboration, the countries with the highest MCP in absolute terms are the United States (367), China (373), and the United Kingdom (175). In terms of the proportion of international collaboration to total production, the country with the highest proportion of international collaboration is the United Kingdom (42.50%), followed by Australia (37.9%) and Spain (37.1%). Observing the other extreme, the countries with the lowest degree of international collaboration are Norway, the United States, and Canada. The MCP of these two countries is relatively low. It suggests that collaboration within countries is dominant for authors working in these countries.

Figure 4 presents a visual map of cooperation between countries/regions with a temporal layer (2014–2024). The node's size indicates the quantity of documents, and the color indicates the average year of publication in each country separately. The thicker the line connecting two nodes, the higher the cooperation between countries. Furthermore, another analysis element is about centrality: the more centrally the node is positioned, the greater the number of links with other countries.

**Figure 4 – Visualization map of countries with the highest citation in *Energy Economics***

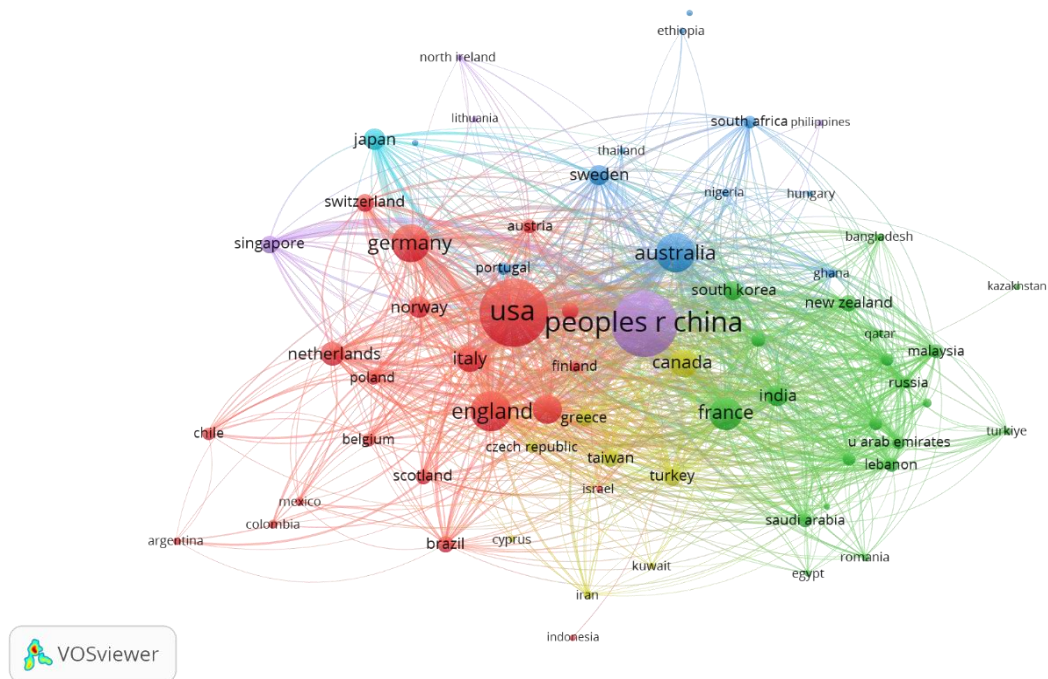

**Source:** Authors' elaboration with data from *Web of Science* and *Scopus*.

Countries such as Malaysia, Saudi Arabia, Pakistan, and Vietnam have relatively younger literature and occupy more peripheral positions in the network visualization map of countries. This peripheral positioning contrasts with the central roles held by China and the United States. Notably, China and the United States exhibit strong ties in scientific cooperation, as depicted in the map. However, a significant distinction lies in the average age of their literature. While the average age of American literature hovers around 2014, Chinese contributions are around 2020. This observation is not surprising, considering China also leads in RD&D investments in renewable energies, as evidenced in the 2021 rankings (Bhutada, 2022).

Figure 5 illustrates the annual publication counts of the top 5 countries contributing to the total publications in *Energy Economics*. From 2009 onwards, these countries experienced an upward trend in scientific output directed towards the journal, a trend likely bolstered by the utilization of the World Wide Web for dissemination and submission purposes. However, after 2017, there was a decline in scientific output from the United States, which was only reversed in 2021.

**Figure 5 – Top 5 countries with the highest participation in the number of publications in *Energy Economics***

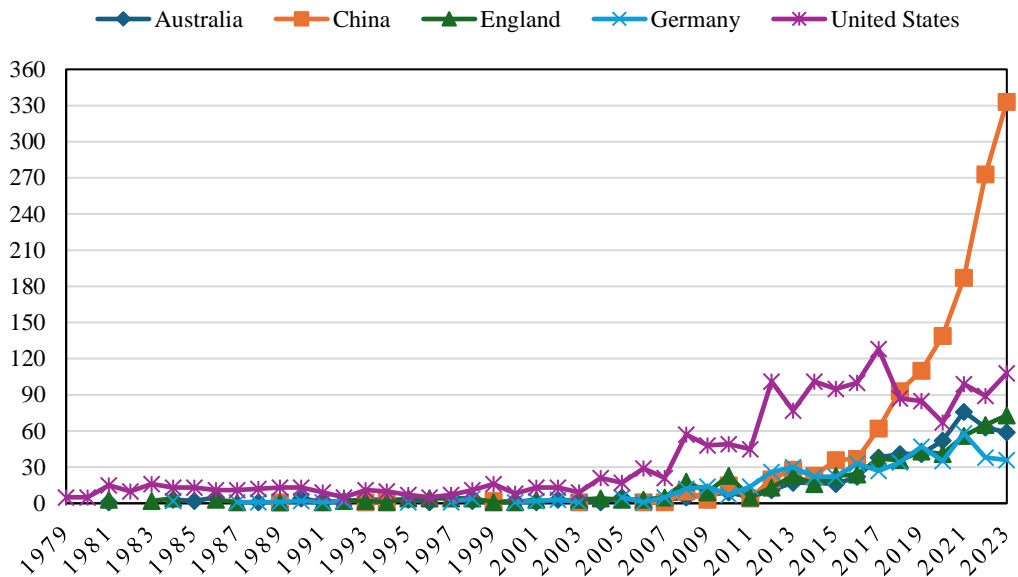

**Source:** Authors' elaboration with data from *Web of Science* and *Scopus*.

In particular, the United States exhibits the highest number of publications between 1997 and 2018, when it matches China's publication count. Following this period, there is a notable surge in Chinese scientific production. China's leadership from 2017 onward is also evident in other specialized journals, as Yao et al. (2023) demonstrated. The United Kingdom has consistently contributed to the journal since its inception. Australia, Germany, and the United Kingdom displayed similar trajectories of scientific production from 2007 to 2020. In 2021, England surpassed Germany, which shows a slight decline from this year onward. Meanwhile, Australia emerges as the third country with the most publications.

### 4.3. Most Productive and Influential Institutions in *Energy Economics*

Table 5 showcases the top ten most productive and influential institutions and their respective countries. Like the computation for countries, the algorithm captures data on institutions from the affiliations registered in author information. In terms of total citations (TC), the National University of Singapore leads the list of the most influential institutions with 6,902 citations, followed by Xiamen University (6,761 citations), the Chinese Academy of Sciences (6,276 citations), and Beijing Institute of Technology with (6,152 citations). Regarding productivity measured by total publications (TP), Xiamen University continues to lead the list, contributing 119 articles. The Chinese Academy of Sciences ranks second with 98 publications, followed by Beijing Institute of Technology (94 publications), Southwestern University of Finance and Economics, and Monash University (both institutions with 89 publications). Collectively, these constitute the five most productive institutions in Energy Economics.

**Table 5 – Most Productive and Influential Institutions in *Energy Economics* between 1979 and 2024**

| Institution | Country | TC | AC | TP |
|---|---|---|---|---|
| Xiamen University | China | 6761 | 56,81 | 119 |
| Chinese Academy of Sciences | China | 6276 | 64,04 | 98 |
| Beijing Institute of Technology | China | 6152 | 65,44 | 94 |
| Southwestern University of Finance and Economics | China | 2952 | 33,17 | 89 |
| Monash University | Australia | 4798 | 53,91 | 89 |
| University of Cambridge | England | 3816 | 43,36 | 88 |
| National University of Singapore | Singapore | 6902 | 85,20 | 81 |
| University of International Business and Economics | China | 3696 | 72,47 | 51 |
| Massachusetts Institute of Technology | United States | 2321 | 33,15 | 70 |
| Lebanese American University | United States | 1076 | 17,35 | 62 |
| RMIT University | Australia | 3006 | 49,27 | 61 |
| Central South University | China | 2665 | 45,94 | 58 |
| Renmin University | China | 1112 | 19,85 | 56 |

**Source:** Authors' elaboration with data from *Web of Science* and *Scopus*.

As depicted in Table 5, the primary institutions of scientific production in the *Energy Economics* collection are universities in China: Xiamen University, Chinese Academy of Sciences, and Beijing Institute of Technology. Out of the top ten universities, seven are Chinese. The University of Cambridge is the sole UK institution on the list, with 88 publications and 3,816 citations. Both Australian universities, Monash University and RMIT University, exhibit significant average citation rates per document (53.91 and 49.27). The National University of Singapore stands out as the institution with the highest citation rate per document in the ranking (85.20). The North American institutions Massachusetts Institute of Technology and Lebanese American University are also featured in the top ten universities with the highest production and influence in the journal *Energy Economics*.

### 4.4. Most Cited Journals in *Energy Economics*

When at least one article from two distinct journals is cited simultaneously in an article published in *Energy Economics*, both journals will thus have a co-citation relationship. This analysis enables us to identify which journals are central to a particular research field – particularly the energy sector. Table 6 lists the top ten most cited journals in *Energy Economics* between 1979 and 2024.

**Table 6 – Most Co-cited Journals in *Energy Economics* between 1979 and 2024**

| Source title | TC | IF 2022 |
|---|---|---|
| *Energy Economics* | 663 | 12,8 |
| Energy Policy | 279 | 9,0 |
| American Economics Review | 168 | 10,7 |
| Econometrica | 90 | 6,1 |
| Review of Economics and Statistics | 81 | 8,0 |
| Energy | 80 | 8,9 |
| The Energy Journal (ENERG J) | 63 | 2,9 |
| Journal of Financial Economics | 63 | 8,9 |
| Journal of Environmental Economics and Management | 60 | 4,6 |
| Journal of Political Economy | 56 | 8,2 |
| Journal of Finance | 47 | 8,0 |
| Review of Economic Studies | 47 | 5,8 |
| Renewable and Sustainable Energy Reviews | 46 | 15,9 |

**Source:** Authors' elaboration with data from *Web of Science* and *Scopus*.

Topping the co-citation ranking is *Energy Economics* itself. Energy Policy is the most cited journal in Energy Policy, with 663 co-citations, followed by Energy Policy, cited 279 times, and the American Economic Review, with 168 co-citations. *Energy Economics* is the journal with the highest number of co-citations, which is unsurprising as journals often engage in self-citation. It is often motivated by consolidating a research line developed by the editorial board over time. Furthermore, the most frequently cited journals in *Energy Economics* are journals with high impact factors in their respective disciplines, such as energy economics, econometrics, financial economics, and environmental economics.

## 4.5. Keyword Co-occurrence Analysis

Keywords are words provided by the author or extracted from the text (title, abstract, or body of the document) seeking to synthesize the main content of the research in terms of hypotheses, methods, and/or evidence. Thus, analyzing the co-occurrence of specific keywords among documents in a research field can indicate a possible predominant theoretical and empirical approach in the literature and possible trends and frontiers to be developed.

This analysis allows the frequency with which keywords appear among the compiled publications to be observed and their co-occurrence with other themes and concepts. When temporal overlay is applied to the visualization map, it further highlights research and publication trends. Therefore, the analysis will initially be applied by integrating all available years in the sample (1979-2024). Then, to provide deeper insights and a more detailed historical description, the sample period will be divided using the periods in which an editor was in charge of the editorial board as a criterion.

## 4.6. Thematic Trends in *Energy Economics* (1979 – 2024)

Regarding the overall keyword co-occurrence analysis, Figure 6 displays the connections between keywords used by authors throughout the journal's activity. The text and size of the points reflect the frequency of the keywords in the records. The width of the lines between the points represents the frequency of co-occurrence of the two keywords.

**Figura 6 – Visualization Map of Keyword Co-occurrence in *Energy Economics* Publications**

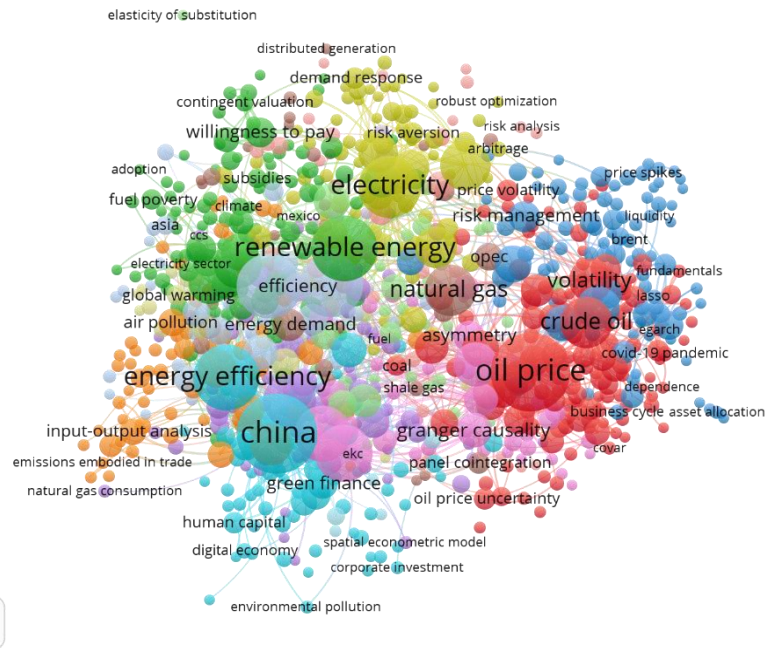

**Source:** Authors' elaboration with data from *Web of Science* and *Scopus*.

Figure 6 shows that themes related to the oil market and its impacts on international price levels have a consolidated space in *Energy Economics*, grouped in the red cluster. The most cited articles in this segment are Sadorsky (1999), Sadorsky (2009), and Sadorsky (2012). Renewable energy transition, energy sources, energy consumption, efficiency, and intensity are also major themes addressed. Apergis & Payne (2012) made significant contributions to the literature on the nexus between energy consumption and economic growth, analyzing heterogeneous effects between renewable and non-renewable energy. Meanwhile, Menegaki (2011) provided evidence suggesting neutrality in the nexus above. Hirth (2013) analyzed the solar and wind energy markets, aiming to develop a better understanding of how their market value is affected by generation variability.

The topic of renewable energies is closely related to electricity markets, another extensively covered theme in *Energy Economics* publications. Zugno et al. (2013) modeled a game that considers the Stackelberg relationship between retailers (leaders) and consumers (followers) in a dynamic pricing environment for the electricity market, providing relevant evidence for the sector. Additionally, in all clusters, econometric techniques are analyzed through various approaches. Particularly, analyses concerning consumer behavior towards energy consumption decisions are assigned a well-defined cluster, presented in light blue. Table 7 below lists the top 20 keywords most cited by authors in the journal *Energy Economics* from 1979 to 2024.

**Table 7 – Top 20 Keywords with the Highest Occurrences in *Energy Economics***

|   | Keyword | Occurrences | Total link strength |
|---|---|---|---|
| 1 | china | 269 | 102 |
| 2 | oil price | 248 | 98 |
| 3 | carbon dioxide emissions | 243 | 95 |
| 4 | renewable energy | 226 | 88 |
| 5 | energy efficiency | 173 | 73 |
| 6 | energy | 162 | 91 |
| 7 | electricity | 161 | 75 |

| 8 | climate change | 158 | 77 |
| 9 | electricity markets | 153 | 60 |
| 10 | energy consumption | 141 | 69 |
| 11 | economic growth | 134 | 72 |
| 12 | natural gas | 105 | 45 |
| 13 | crude oil | 104 | 57 |
| 14 | climate policy | 95 | 47 |
| 15 | energy poverty | 79 | 19 |
| 16 | forecasting | 76 | 57 |
| 17 | oil price shocks | 73 | 39 |
| 18 | volatility | 73 | 50 |
| 19 | energy intensity | 71 | 35 |
| 20 | data envelopment analysis | 70 | 34 |

**Source:** Authors' elaboration with data from *Web of Science* and *Scopus*.

It is important to note that when keywords are grouped, their importance over time may be overlooked. For this reason, in Figure 7, we present the top 10 keywords by occurrence in *Energy Economics* in different periods of its history. Naturally, fluctuations in each keyword are partly explained by the rise of specific issues, technological progress, and the maturation of particular fields of exploration – such as climate change and the quest for its mitigation effects. We see in Figure 7 that the topics "renewable energy" and "carbon dioxide emissions" have grown in the last 20 years. The decrease in the use of the keywords "oil price" and "energy consumption" contrasts with the resurgence in the use of "energy efficiency."

**Figure 7 – Top 10 words with the highest number of occurrences in *Energy Economics***

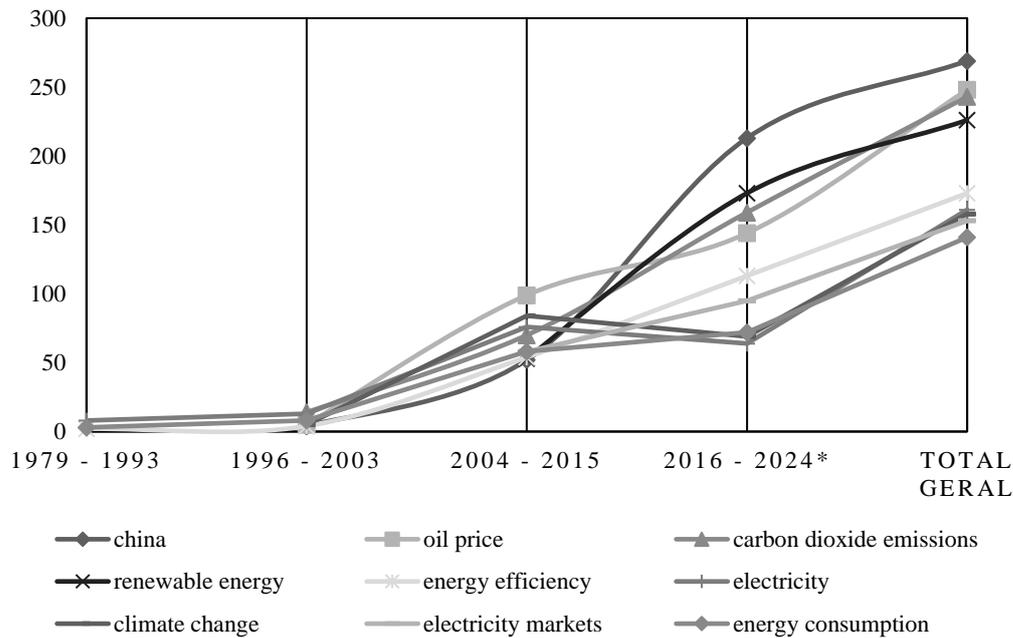

**Source:** Authors' elaboration with data from *Web of Science* and *Scopus*.

The research on China experienced a steep increase from 2004 onwards, becoming the keyword with the highest number of occurrences throughout the journal's history. Articles related to "oil prices," "climate change," and "electricity" show a slight decline between the periods 2004-2015 and 2016-2024. The reduction in "electricity," in particular, contrasts with the increase in the occurrence of "electricity markets" (Kara et al., 2008; Kumbaroğlu et al., 2008). Figure 8 presents the eleventh to twentieth keywords with the highest occurrence in *Energy Economics*. We see that the theme of economic growth stands out in all periods.

The use of the term "*climate policy*" shows a slight decline between the periods 2004-2015 and 2016-2024, along with the decline of "*data envelopment analysis*," "*energy intensity*," and "*volatility*." Particularly notable is the prominence of the concept of "*energy poverty*" in the last period, which has become one of the major concerns on the agenda of public policies in many countries (UN, 2015a).

**Figure 8 – Eleventh to twentieth most frequently occurring keyword in *Energy Economics***

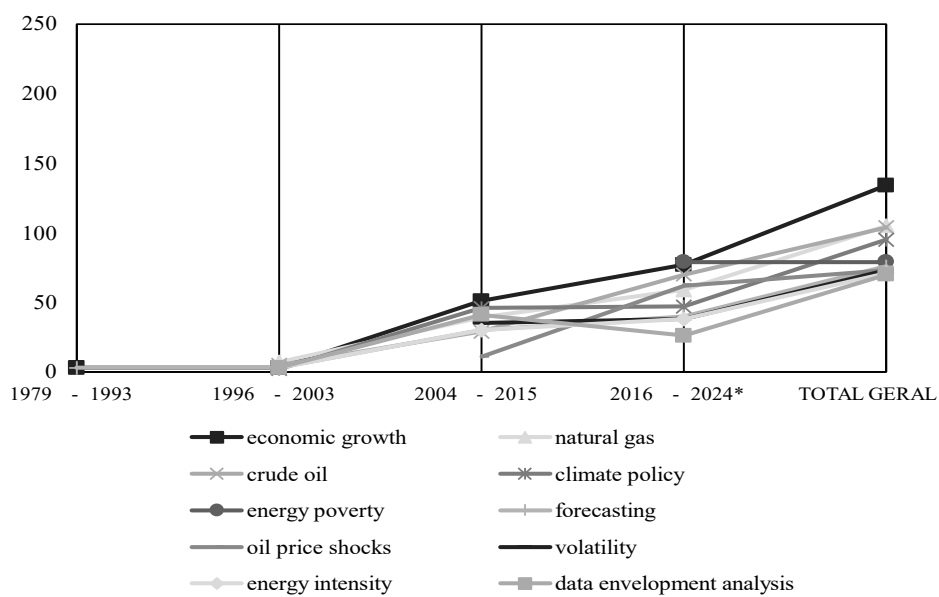

**Source:** Authors' elaboration with data from *Web of Science* and *Scopus*.

These studies aim to provide empirical evidence on the phenomenon using econometric methods and microdata (Awaworyi Churchill & Smyth, 2020, 2021; Koomson & Danquah, 2021; Zhao et al., 2021).

### *4.7. Analysis by Editorial Periods*

In this subsection, we will delve into the scientific evolution of *Energy Economics* across different editorial periods.

### *4.7.1. First Stage: 1979 – 1993*

During the initial stage of the journal (1979–1993), *Energy Economics* was under the editorial leadership of Homa Motamen-Scobie, with Colin Robinson's contribution in the first four years. A co-occurrence analysis of keywords was conducted for this period. Figure 9 depicts the keyword connection map used during this timeframe. Notably, few publications contributed to this analysis - evidenced by the limited number of nodes. Nevertheless, the results allow us to observe the main topics addressed right from the inception of this technical-scientific discussion space.

Between 1979 and 1993, the energy sector underwent significant global transformations and challenges driven by geopolitical events, technological advancements, and changes in energy policies. The 1979 oil crisis profoundly impacted global energy security, leading many countries to reassess their reliance on fossil fuels and seek alternative and renewable sources. Simultaneously, the nuclear race and the search for more sustainable energy production methods were recurring themes, reflecting environmental concerns and energy security.

**Figure 9 – Keyword network in *Energy Economics* in editions from 1979 to 1993**

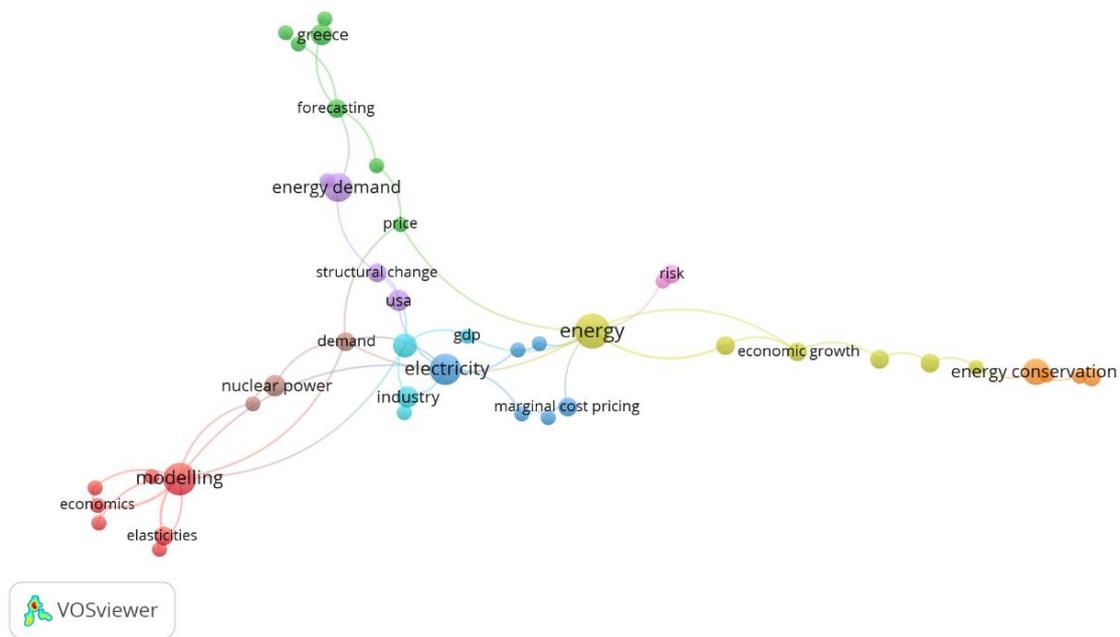

**Source:** Authors' elaboration with data from *Web of Science* and *Scopus*.

We observe that fundamental concepts such as "elasticities," "economic growth," "demand," and "marginal cost pricing" are relevant in their respective clusters, indicating rather pure theoretical and empirical approaches on the topics, which were declared objectives of *Energy Economics* in its first editorial letter (Motamen & Robinson, 1979). During this period, with more data availability (especially time series), several macroeconometric applications sought to support economic analyses and public policy for the energy sector. It is evidenced by the presence of "structural change" associated with "energy demand" (Ang & Lee, 1994; Jenne & Cattell, 1983). Additionally, a cluster is centered around the keyword "nuclear power," indicating that many discussions in the energy sector revolved around using economic theory to implement this type of source to meet the growing energy demand (Pearce, 1979; Sutherland, 1986).

The red cluster is centrally linked to the keyword "modeling." In the energy sector, particularly, economic modeling aids decision-making and strategic planning, enabling more efficient resource allocation. For example, the works of Proops (1984) and Ang (1987) investigate the relationship between GDP and energy consumption. In the oil sector, Dailami (1983) uses time series modeling to analyze short-term oil demand behavior – a shared objective with Kuenne (1982), who addresses the GENESYS model of the Organization of the Petroleum Exporting Countries (OPEC). In the electricity supply sector, demand modeling (residential, commercial, and industrial) was conducted (Bolle, 1992; Hankinson & Rhys, 1983; Leung & Hsu, 1984) and also subject to methodological scrutiny – especially considering significant structural breaks such as the 1973 oil embargo (Sutherland, 1986).

*4.7.2. Second Stage: 1996 – 2003*

Between 1996 and 2003, Derek Bunn led the editorial work of *Energy Economics*, with the collaboration of Musa Essayyad and Richard Tol, between 2002 and 2003. In this subsection, keyword co-occurrence will be analyzed from 1996 to 2003.

The energy sector experienced significant global transformations and challenges during this period, significantly impacting energy policies, technologies, and economies worldwide. The period was characterized by the opening of energy markets, aiming to promote competition and the entry of new players into the sector (Hauch, 2003; Huisman & Mahieu, 2003). The search for renewable energy sources gained prominence, driven by environmental concerns and the growing awareness of climate change (Hoekstra & van den Bergh, 2003; van der Zwaan et al., 2002). Similarly, the level of carbon dioxide emissions is represented in the dark blue cluster, connected to keywords such as "market structure," "coal," and "carbon taxes" (Fisher-Vanden et al., 1997; Greening et al., 1998). In 1996, the special edition "Energy-Environmental Modeling," edited by D. Bunn, E. Larsen, and K. Vlahos, showcased articles from the 1995 Symposium on Energy Models for Policy and Planning, held at the London Business School in association with the International Federation of Operational Research Societies. The motivation behind the symposium was to focus on model-based insights on energy issues (Bunn et al., 1997).

In Figure 10 we present the keyword network for the respective period.

**Figura 10 – Keyword network in *Energy Economics* in editions from 1996 to 2003**

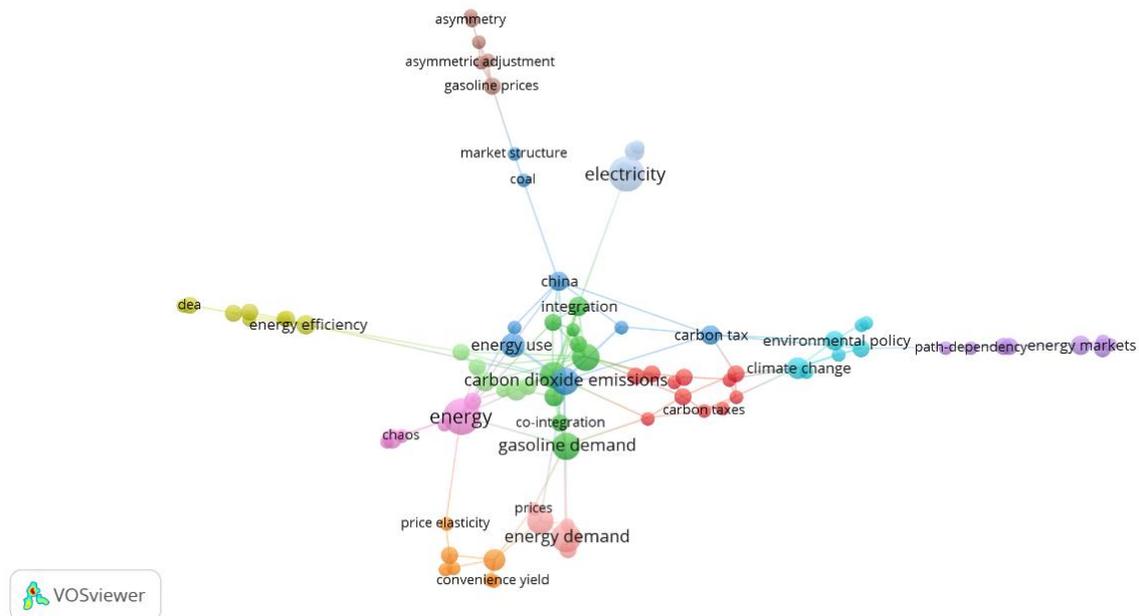

**Source:** Authors' elaboration with data from *Web of Science* and *Scopus*.

Furthermore, the impacts of climate change are represented by the light blue cluster, centered by the keywords "climate change" and "environmental policy" (Carraro & Hourcade, 1998; Mabey & Nixon, 1997). There is also a concern with the fuel market, represented by the node "gasoline demand" in the green cluster and "gasoline prices" in the brown cluster, well above in Figure 10 (Espey, 1998; Galeotti et al., 2003).

*4.7.3. Third Stage: 2004 – 2015*

From 2004 to 2015, the energy sector reflected increasing concerns about energy security, environmental sustainability, and the search for alternative energy sources. During this period, *Energy Economics* had various editors and co-editors. From 2004 to 2007, Richard Tol and John Weyant led the journal's edition. Derek Bunn also held the position of honorary editor during this time. From 2008 to 2013, Beng W. joined the editorial coordination alongside the previously mentioned editors. Ugur Soytas joined their efforts between 2014 and 2015. They are renowned academics and researchers in energy economics, climate economics, evaluation of energy policies, and analysis of the economic impacts of climate change.

With the increasing complexity of energy markets and their association with global phenomena such as climate change, there was a stimulus to create economic instruments for reducing carbon emissions, promoting transition, and energy efficiency. Significant investments in clean energy sources such as solar, wind, and biofuels led to technological advancements and the implementation of incentive policies in various countries. In light of this, Table 9 lists the regular and special editions produced by the journal during this period.

Table 8 – Issues and special issues (2004 – 2015)

| Year | Title | Editors | DOI |
|---|---|---|---|
| 2004 | EMF 19 Alternative technology strategies for climate change policy | John P. Weyant | 10.1016/j.eneco.2004.04.019 |
| 2005 | Special Issue on Electricity Markets | Derek Bunn | (*) |
| 2006 | Modeling Technological Change in Climate Policy Analyses | John Houghton | 10.1016/j.eneco.2006.05.010 |
| 2007 | Modeling of Industrial Energy Consumption | Lorna A. Greening; Gale Boyd; Joseph M. Roop | 10.1016/j.eneco.2007.02.011 |
| 2008 | Technological Change and the Environment | Karen Fisher-Vanden | 10.1016/j.eneco.2008.08.001 |
| 2009 | Technological Change and Uncertainty in Environmental Economics | Christoph Böhringer; Tim P. Mennel; Tom F. Rutherford | 10.1016/j.eneco.2009.05.006 |
| 2009 | Energy Sector Pricing and Macroeconomic Dynamics | Catherine Kyrtsou; Anastasios G. Malliaris | 10.1016/j.eneco.2009.08.019 |
| 2009 | International, U.S. and E.U. Climate Change Control Scenarios: Results from EMF 22 | Leon Clarke; Christoph Böhringer; Tom F. Rutherford | 10.1016/j.eneco.2009.10.014 |
| 2010 | Policymaking Benefits and Limitations from Using Financial Methods and Modelling in Electricity Markets | Richard Green; Benjamin Hobbs; Shmuel Oren; Afzal Siddiqui | 10.1016/j.eneco.2010.04.012 |
| 2011 | Special Issue on The Economics of Technologies to Combat | Nebojsa Nakicenovic; William | 10.1016/j.eneco.2011.02.001 |

| Year | Title | Authors | DOI |
|---|---|---|---|
| | Global Warming | Nordhaus | |
| 2011 | Fourth Atlantic Workshop in Energy and Environmental Economics | Carlos de Miguel; Xavier Labandeira; Baltasar Manzano | 10.1016/j.eneco.2011.07.019 |
| 2012 | Green Perspectives | Brian P. Flannery; Richard S.J. Tol | 10.1016/j.eneco.2012.09.006 |
| 2012 | The Asia Modeling Exercise: Exploring the Role of Asia in Mitigating Climate Change | Katherine Calvin; Leon Clarke; Volker Krey | 10.1016/j.eneco.2012.09.003 |
| 2012 | The Role of Border Carbon Adjustment in Unilateral Climate Policy: Results from EMF 29 | Christoph Böhringer; Edward J. Balistreri; Thomas F. Rutherford | 10.1016/j.eneco.2012.10.002 |
| 2013 | Quantitative Analysis of Energy Markets | Angelica Gianfreda; Luigi Grossi | 10.1016/j.eneco.2012.06.026 |
| 2013 | Fifth Atlantic Workshop in Energy and Environmental Economics | Carlos de Miguel; Alberto Gago; Xavier Labandeira; Baltasar Manzano | 10.1016/j.eneco.2013.10.002 |
| 2014 | Special Issue on Recent Approaches to Modelling Oil and Energy Commodity Prices | Matteo Manera | 10.1016/S0140-9883(14)00313-2 |
| 2014 | Climate Adaptation: Improving the connection between empirical research and integrated assessment models | Karen Fisher-Vanden; David Popp; Ian Sue Wing | 10.1016/j.eneco.2014.11.010 |
| 2015 | Frontiers in the Economics of Energy Efficiency | Carlos de Miguel; Xavier Labandeira; Andreas Löschel | 10.1016/j.eneco.2015.11.012 |

**Source:** Authors' elaboration. (*) There is no document introducing this special issue with a DOI available. This edition can be found in volume 27, issue 2 (2005).

Studies on technology, climate change, and economics in the energy sector were encouraged in the journal. In 2004, with an emphasis on alternative technology strategies for climate change policies, special editions addressed topics such as electricity markets, industrial energy consumption, and environmental technological changes. Additionally, there was a growing interest in the relationship between technology, uncertainty, and environmental economics, as well as modeling international climate policies and the macroeconomic effects of the energy sector. Special editions also explored innovative approaches to combat global warming, focusing on the importance of energy efficiency. These themes reflect the energy sector's historical changes and challenges, demonstrating the constant search for more sustainable and effective solutions in this field.

Figure 11 shows the keyword network for the period from 2004 to 2015.

**Figure 11 – Keyword network in *Energy Economics* in editions from 2004 to 2015**

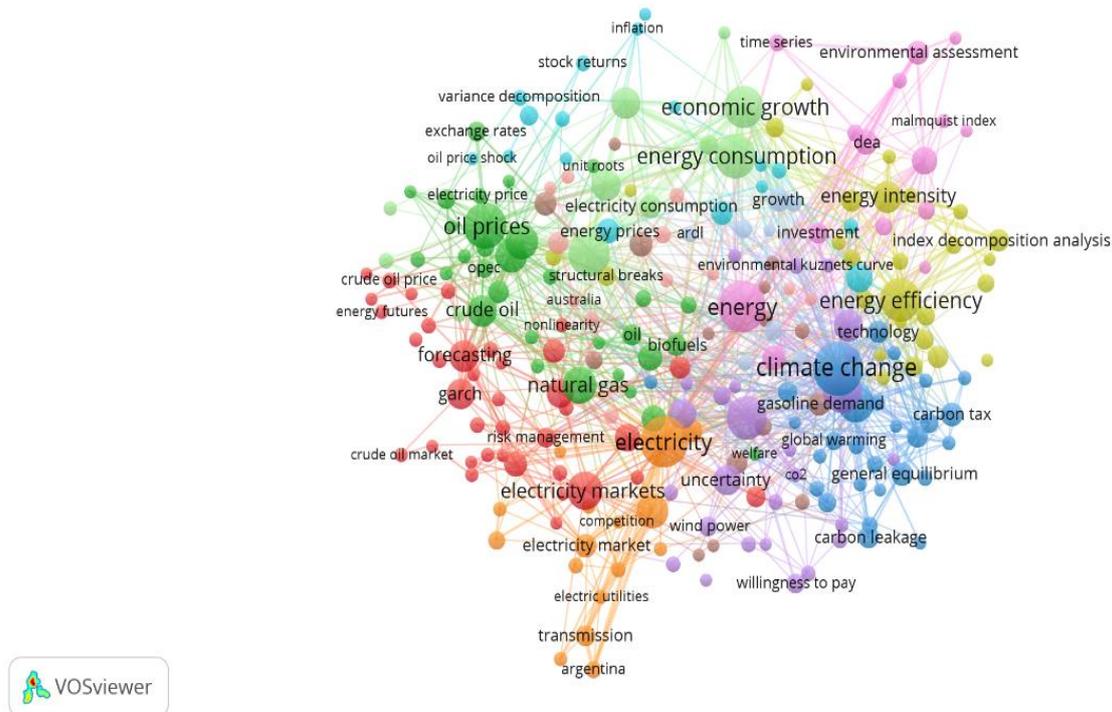

**Source:** Authors' elaboration with data from *Web of Science* and *Scopus*.

Thus, the energy sector benefited from theoretical and interpretative insights and cost-benefit analyses, economic evaluations, and policy impact assessments. The purple cluster, located in the bottom right of Figure 11, features keywords such as "*willingness to pay*," "*uncertainty*," "*technology adoption*," and "*renewable energy*," with the latter being the most prevalent (53 publications use it as an identifier for their study). "*Choice experiment*" also emerges close to "willingness to pay," consisting of a robust and widely used method for measuring an individual's propensity to execute (or not) a particular situation or consumption and investment decision (Carlsson & Martinsson, 2008; Sundt & Rehdanz, 2015; Susaeta et al., 2011).

With new regulations allowing private players in the electricity sector in various countries, "electricity markets" emerges as a highly centralized keyword in the red cluster. It is related to other keywords such as "regulation," "market power," and "competitiveness." During the 1980s and 1990s, there was a significant movement towards liberalizing electricity markets in various countries worldwide. The process introduces more competition and efficiency into the electricity sector, moving away from monopolistic and state models. The opening of these markets allowed the participation of new players, such as private companies and foreign investors, encouraging innovation, reducing costs, and promoting operational efficiency. This discussion is brought by Willems et al. (2009) and (van der Veen et al., 2012).

Furthermore, the growth of the cluster related to the keyword "energy efficiency," positioning itself close to the cluster related to climate change (dark blue) and the cluster related to renewable energies (purple), is noteworthy. Banfi et al. (2008) and Brounen et al. (2013) highlight this topic in the period.

*4.7.4. Fourth Stage: 2016 to Present*

This stage spans from 2016 to April 2024. During this period, Richard Tol served as the editor-in-chief of *Energy Economics*, with the collaboration of John Weyant as honorary editor. John Weyant and Richard Tol had already served as editors in previous journal stages, either as editors or associate editors.

Table 11 presents the regular and special editions of the journal *Energy Economics* published during this period. In 2015, there was a significant focus on energy markets, particularly econometric approaches, and financial and convergence issues in the energy sector. These special editions reflect the increasing complexity and interconnection of the energy sector's environmental, financial, and geopolitical aspects during this period.

**Table 9 – Editorial Letters and Special Issues (2016 – 2024)**

| Year | Title | Editors | DOI |
|---|---|---|---|
| 2015 | Econometrics of Energy Markets | Alfred Deakin | 10.1016/j.eneco.2015.07.001 |
| 2016 | Energy Markets | Rita Laura D'Ecclesia | 10.1016/j.eneco.2015.11.014 |
| 2016 | Special Section on the findings of the CLIMACAP-LAMP project | Katherine V Calvin; B.C.C. van der Bob Zwaan; Leon E Clarke | 10.1016/j.eneco.2016.05.005 |
| 2017 | Seventh Atlantic Workshop in Energy and Environmental Economics | Carlos de Miguel; Massimo Filippini; Xavier Labandeira; Andreas Löschel | 10.1016/j.eneco.2018.01.031 |
| 2017 | Energy Finance and Energy Markets | Paresh Kumar Narayan | 10.1016/j.eneco.2017.08.021 |
| 2017 | Special Issue on Energy Sector Convergence | Nicholas Apergis; James E. Payne; Bradley T. Ewing | 10.1016/j.eneco.2016.09.015 |
| 2017 | Energy, Commodities and Geopolitics: Modeling Issues | Anna Creti; Marc Joets; Matteo Manera; Lutz Kilian | 10.1016/j.eneco.2016.08.022 |
| 2017 | Variable Renewable Electricity and Power Sector Dynamics in Integrated Assessment Models | Douglas Arent; Gunnar Luderer | 10.1016/j.eneco.2017.03.027 |
| 2018 | The EMF 32 study on technology and climate policy strategies for greenhouse gas reductions in the U.S. electric power sector | John Bistline; Jared Creason; Brian Murray | 10.1016/j.eneco.2018.03.007 |
| 2019 | Energy Finance 2017 | Paresh Kumar Narayan | 10.1016/j.eneco.2018.12.001 |

**Source:** Authors' elaboration.

Amidst a growing need for decarbonizing the energy matrix, the significant increase in investment in renewable energies such as solar, wind, and biomass is depicted by the green cluster, a node with high centrality in Figure 12. The transition to a low-carbon economy has become a global priority, resulting in stricter regulations, emissions reduction targets, and incentives for adopting more sustainable practices (Adebayo et al., 2021; Aller et al., 2021; Bjelle et al., 2021). As shown in Figure 12, the topic of energy economics becomes extensive, encompassing various dimensions in its fundamental nature as an applied social science.

**Figure 12 – Keyword network in *Energy Economics* in editions from 2016 to 2024**

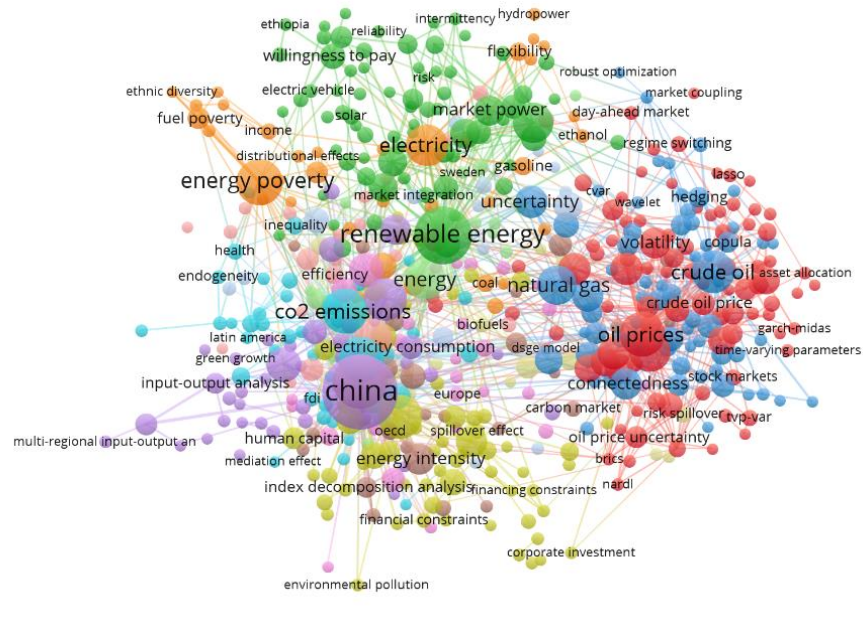

**Source:** Authors' elaboration with data from *Web of Science* and *Scopus*.

Integrating smart technologies, energy storage, and smart grid networks shapes how energy is generated, distributed, and consumed, paving the way for a more diversified, resilient, and sustainable energy future. At the same time, the same opportunities observed during this period bring along complex social problems that researchers have been delving into to highlight. It is exemplified by the emergence of the cluster centered around "energy poverty." Energy poverty refers to the situation in which individuals or communities lack adequate access to reliable, affordable, and clean energy sources to meet their basic needs (Awaworyi Churchill & Smyth, 2020). Energy poverty can affect the affected populations' quality of life, health, safety, and economic development, creating a cycle of disadvantages and negatively impacting social and environmental well-being. Evidence, such as the association between an increase in energy poverty and a negative perception of individual well-being and health, has been presented by Awaworyi Churchill & Smyth (2021), Awaworyi Churchill et al. (2020), and Pan et al. (2021).

Another noteworthy element of this period is the emergence of the keyword "China" with the high frequency of occurrence and centrality in the overall panorama of term occurrences in the journal. Currently, "China" is quantitatively the most used identifier in the journal's history. Its appearance relates to various distinct topics: "energy consumption" (Ren et al., 2021), "CO2 emissions" (Shahbaz et al., 2020; Zheng et al., 2019), "energy poverty" (Dong et al., 2021; Zhao et al., 2021), and "structural decomposition analysis" (Su & Thomson, 2016).

## 4.8. Future Directions for *Energy Economics*

In this subsection, we primarily investigate articles published in the last four years, namely, 2020 to April 2024. Utilizing thematic evolution maps generated based on the centrality and density metrics, we depict the current state and progression of key topics on a map with four quadrants extracted from the R language package, *bibliometrix* (Aria et al., 2022; Aria & Cuccurullo, 2017). Thematic communities were calculated using the *Walktrap* algorithm, which is highly effective in samples with characteristics similar to this study's (Yang et al., 2016). The groups obtained through community detection can be projected onto a strategic diagram graphical scheme. This diagram organizes the thematic domain under investigation based on two measures known as Callon centrality and Callon density (Aria et al., 2022).

The centrality of a theme refers to its importance in defining the overall development of the research field – in this case, the journal under analysis – while density reflects the growth of that theme. The units of analysis used in this application are the keywords designated by the authors of each document, and the map is constructed based on the co-occurrence of these terms. According to Aria et al. (2022), (i) higher values of centrality and density define hot topics that are well-developed and relevant for structuring the conceptual framework of the domain; (ii) higher values of centrality and lower values of density define basic topics, significant for the domain and transversal to its different areas; (iii) lower values of centrality and density define peripheral topics, not fully developed or marginally interesting for the domain; and (iv) lower values of centrality and higher values of density define niche topics, strongly developed but still marginal for the domain under investigation.

Figure 13 presents the thematic map of *Energy Economics* (2020–2024).

**Figure 13 – Thematic Map of *Energy Economics* (2020–2024)**

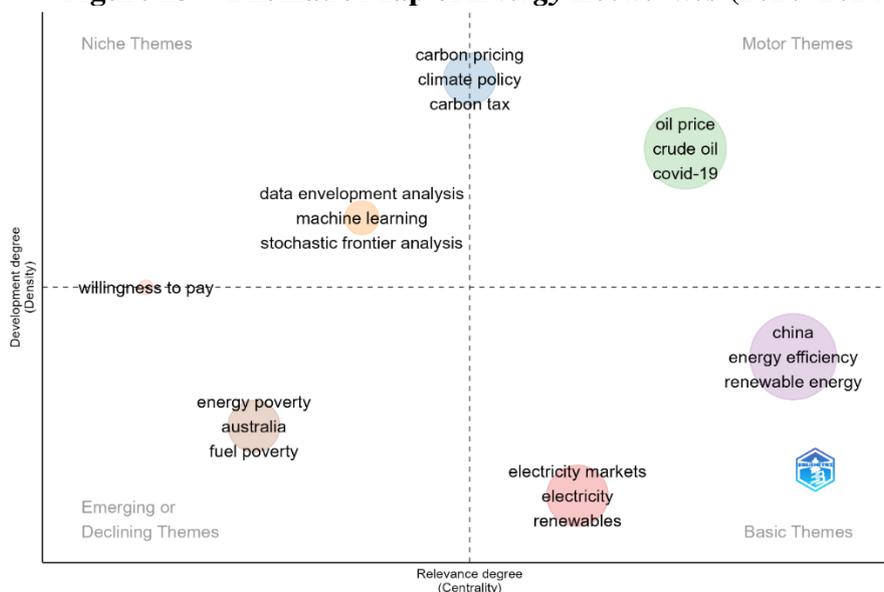

**Source:** Authors' elaboration with data from *Web of Science* and *Scopus*.

Emerging and niche topics indicate that the literature developed in recent years in *Energy Economics* focuses on two main aspects: the development and analysis of economic instruments and climate policies – mainly aiming at reducing CO2 emissions

– and the socio-technical and economic dimensions of energy transition – represented by the concepts of energy poverty and fuel poverty. Both topics have motivated the creation of their sustainable development goals in the UN framework (UN, 2015b) and the creation of a multilateral treaty, such as the Paris Agreement, signed in 2015 by 196 parties (UNFCCC, 2021).

Moreover, the agreement emphasizes the need for financial and technological support for developing countries to deal with the impacts of climate change. The energy transition, by seeking to shift from more polluting energy sources to cleaner and more sustainable ones, can inadvertently contribute to the increase in energy poverty (Koomson & Danquah, 2021; Nguyen & Nasir, 2021). Implementing greener technologies increases consumer costs, making energy more inaccessible to low-income populations. Additionally, the necessary infrastructure for renewable energies may not be adequately distributed, leaving deprived communities disconnected or with limited access to these sources (Dong et al., 2021). Therefore, it is crucial to consider measures and policies that ensure the energy transition is inclusive and equitable, avoiding exacerbating energy poverty.

The analysis of willingness to pay plays a fundamental role in climate change adaptation, as it allows people to assess their willingness to invest in measures and technologies that can mitigate environmental impacts and promote resilient communities (Sundt & Rehdanz, 2015). Understanding preferences and how much people are willing to spend to protect the environment, adopt certain technologies, and adapt to climate change is crucial for developing effective strategies and formulating economically viable and socially acceptable policies and initiatives. For this reason, we see that "willingness to pay" occupies a particular position on the thematic map of *Energy Economics*, situated among both niche literature and emerging topics of the journal. It is worth noting that applying contingent valuation methods requires considerable investment in time, resources, and expertise and a careful approach to dealing with associated methodological challenges, representing a significant asset to the journal. DEA is a non-parametric method for measuring productivity and frontier production generally does the same with parametric methods. This topic is presented in the graph as a niche theme, with intermediate values of centrality and density value.

Based on the preceding discussion, we believe that the journal Energy Economics is well positioned to address the challenges and opportunities of the international energy sector, engaging in robust discussions critical to promoting countries' social and economic well-being across various development contexts.

## 5. Final Remarks

Based on data from the *Web of Science* and *Scopus* databases, this article comprehensively analyzes the history and future trends of the journal *Energy Economics*. Using bibliometric methods, including performance analysis and scientific mapping tools, we have identified the main themes elaborated by the journal, situating it in the context of the energy sector from 1979 to 2024. *Energy Economics* is now one of the leading journals in energy economics with a high impact factor, providing many seminal insights into the oil sector and its interactions with stock and commodity markets. Its robust internationalization allows it to provide evidence on issues related to the sector, considering different contexts and applications.

The results show that China leads the number of publications in *Energy Economics*, and the country is widely studied – as evidenced by keyword co-occurrence analysis and the scientific performance of countries and institutions – notably influential and productive in this field of exploration. Over the years, *Energy Economics* has been rigorous in its editorial policies, ensuring that topics that motivated the journal's inception – the dynamics of the oil sector – still constitute the driving themes of its scientific contribution today. This rigor also applies to the reading and approach to new topics in its collections.

Despite the increasing complexity of co-occurrence networks over editorial phases, we see that over the past decade, only one cluster has been consistently included – "energy poverty" – among the major themes addressed by the journal. This topic is crucial to social development, stemming from the multidimensionality that the field of energy economics demands. In addition, the intersection with the issue of climate change and renewable energies has become consistently present in the empirical analyses of *Energy Economics*, including the impact of climate change, technological innovations and diffusion, social equity, government initiatives, and consumer preferences.

Some limitations in this study inherent to applying bibliometric techniques should be noted. Despite our efforts to provide context for keyword co-occurrence analysis, there are still opportunities to explore the themes and concepts in the maps and tables further. Thus, caution should be exercised regarding overestimating or underestimating the importance of certain concepts. Therefore, as future research, we suggest a detailed analysis of clustered topics. Furthermore, this study reflects the thematic development of the journal based on the available periods. Its update may be required without compromising knowledge accumulation. This article proposes a systematic approach to the main bibliographic metrics. It will guide the reader in conducting analyses in fractal layers based on their particular interest with greater agility and criteria.